\documentstyle[epsfig,aps,prb]{revtex}
\newcommand\beq{\begin{equation}}
\newcommand\eeq{\end{equation}}
\newcommand\bea{\begin{eqnarray}}
\newcommand\eea{\end{eqnarray}}
\newcommand\ed{E /\delta }
\newcommand\qd{q /\delta }
\newcommand\psr{\psi_R }
\newcommand\psl{\psi_L }
\newcommand\psp{\psi_+ }
\newcommand\psm{\psi_- }
\newcommand\vd{V_{\delta} }

\begin{document}

\begin{center}
{\Large One-dimensional fermions with incommensuration}
\end{center}

\vskip .5 true cm
\centerline{\bf Diptiman Sen \footnote{E-mail address: 
diptiman@cts.iisc.ernet.in} and Siddhartha Lal \footnote{E-mail address:
sanjayl@cts.iisc.ernet.in}}
\vskip .5 true cm

\centerline{\it Centre for Theoretical Studies, Indian Institute of Science,}  
\centerline{\it Bangalore 560012, India} 
\vskip .5 true cm

\begin{abstract}

We study the spectrum of fermions hopping on a chain with a weak 
incommensuration close to dimerization; both $q$, the deviation of the wave 
number from $\pi$, and $\delta$, the strength of the incommensuration, are 
small. For free fermions, we use a continuum Dirac theory to show that there 
are an infinite number of bands which meet at zero energy as $q$ approaches 
zero. In the limit that the ratio $q/ \delta \rightarrow 0$, the number of 
states lying inside the $q=0$ gap is nonzero and equal to $2 \delta /\pi^2$. 
Thus the limit $q \rightarrow 0$ differs from $q=0$; this can be seen clearly 
in the behavior of the specific heat at low temperature. For interacting 
fermions or the $XXZ$ spin-$1/2$ chain close to dimerization, we use 
bosonization to argue that similar results hold; as $q \rightarrow 0$, we find 
a nontrivial density of states near zero energy. However, the limit $q 
\rightarrow 0$ and $q=0$ give the same results near commensurate wave numbers 
which are {\it different} from $\pi$. We apply our results to the 
Azbel-Hofstadter problem of electrons hopping on a two-dimensional lattice in 
the presence of a magnetic field. Finally, we discuss the complete energy
spectrum of noninteracting fermions with incommensurate hopping by
going up to higher orders in $\delta$.

\end{abstract}
\vskip 1 true cm

~~~~~~~~~~~ PACS number: ~71.10.Fd, ~71.10.Pm, ~75.10.Jm
\vskip 1 true cm

\newpage

\section{Introduction}

One-dimensional lattice models with incommensurate hopping elements or 
on-site potentials have been studied for many years from different points 
of view \cite{sok,val}. Many unusual properties of the quantum
spectra and wave functions have been discovered for various kinds of 
aperiodicity \cite{ost}. Physically, such models have applications to 
several problems such as incommensurate crystals \cite{lyu}, semiconductor 
heterojunctions \cite{mer}, the incommensurate phase of spin-Peierls systems 
such as $CuGeO_3$ \cite{lor}, and the Azbel-Hofstadter problem of particles 
hopping on a two-dimensional lattice in the presence of a magnetic field 
\cite{azb}. If the incommensuration is close to dimerization (wave number 
$\pi$), then the models have an additional interest in the context of 
metal-insulator transitions near half-filling and spin-Peierls systems near 
zero magnetization \cite{bha,nak}.

In this paper, we study a model of fermions with a hopping which has a 
{\it weak} sinusoidal incommensurate term {\it close} to dimerization. We will 
show both numerically and analytically that this has some unusual properties,
particularly inside the gap which exists {\it exactly} at dimerization. The 
major surprise is that states appear inside the gap {\it as soon as} we move 
away from dimerization. These states carry a finite weight, and therefore 
contribute to quantities like the specific heat at low temperatures, as we 
will show. In Secs. II A and II B, we discuss free fermions or the
$XY$ spin-$1/2$ chain close to dimerization (and half-filling) for which 
quantitatively accurate results can be 
obtained by analytical methods. We will use a continuum Dirac field 
formulation, followed by perturbation theory and a WKB analysis, to find the
locations and widths of the energy bands inside the $q=0$ gap; here $q$ is 
the deviation from $\pi$ of the wave number of the incommensuration. In 
Sec. II C, we will explain how the limit $q \rightarrow 0$ get reconciled 
with the results for $q=0$ in a somewhat different model, namely, the 
Azbel-Hofstadter problem in which electrons hop on a two-dimensional 
lattice in the presence of a magnetic field.
In Sec. III, we use bosonization to argue that similar differences between 
$q \rightarrow 0$ and $q=0$ occur for interacting fermions or the $XXZ$ 
spin-$1/2$ chain, which is the more general and interesting case. In Sec. IV,
we show that such peculiarities do not occur in the vicinity of any 
wave number which is {\it different} from $\pi$. In Sec. V, we consider 
continuum theories which go up to higher orders in the strength of the 
incommensuration; this enables us to discuss the complete energy spectrum of
fermions with a weak incommensurate hopping. In Sec. VI, we make some 
concluding remarks.

Our work differs from earlier ones which have concentrated on the effects
of incommensuration which are either large or close to ``highly" irrational
numbers such as the golden ratio \cite{sok,ost,lyu,azb}. Further, we will
use techniques from continuum theory which have not been used much 
in this area (except for the recent work in Ref. 8). 

\section{Noninteracting fermions close to dimerization}

\subsection{Numerical results}

We begin with the following Hamiltonian for noninteracting and spinless 
fermions on a lattice 
\bea
H ~&=&~ - ~\frac{1}{2} ~\sum_n ~J_n ~(~ c_n^\dagger ~c_{n+1} ~+~ 
c_{n+1}^\dagger c_n ~) ~-~ \mu ~\sum_n ~c_n^\dagger ~c_n ~, \nonumber \\
J_n ~&=&~ 1 ~+~ \delta ~\cos (\pi +q) n ~,
\label{ham1}
\eea
where we will assume that $\delta \ll 1$ and $q \ll \pi$. (In that limit,
we will show later that it does not matter if the incommensurate term is put 
in the chemical potential $\mu$ (i.e., on-site) rather than in the hopping 
$J_n$). We set $\mu =0$ as we are interested in energies close to zero. 
If $\delta=0$, the model can be easily solved; the 
dispersion relation is $E(k) = -\cos k$. In the ground state, all the states 
with momenta lying in the range $[-k_F , k_F ]$ are filled, where the Fermi 
momentum $k_F =\pi /2$. The Fermi velocity is equal to $\sin k_F = 1$. 
(We will set both the lattice spacing and Planck's constant equal to $1$).

If $\delta \ne 0$ but $q=0$, i.e., with dimerization, the model can still 
be solved analytically. There is an energy gap extending from $-\delta$ to 
$\delta$. Let us now consider nonzero values of $q$. [Since $\cos (\pi + q)n
= \cos (qn) \cos (\pi n)$, we see that a small value of $q$ is equivalent
to dimerization term whose amplitude $\delta \cos (qn)$ varies slowly with
$n$]. For any rational value of $q/\pi = M/N$, where $M$ and $N$ are 
relatively prime integers, we have a periodic system with period $P$
equal to $N$ if $M+N$ is even and $2N$ is $M+N$ is odd. The one-particle 
spectrum of Eq. (\ref{ham1}) can be found by solving the discrete
Schr\"odinger equation
\beq
- \frac{1}{2} ~[~ J_n ~\psi_{n+1} ~+~ J_{n-1} ~\psi_{n-1} ~] ~=~ E ~\psi_n ~,
\eeq
which is equivalent to the equation
\beq
\Bigl( \begin{array}{l}
\psi_{n+1} \\
\psi_n
\end{array} \Bigr) ~=~ \Bigl( \begin{array}{cc} 
-2E/J_n & - J_{n-1}/J_n \\
1 & 0 
\end{array} \Bigr) ~\Bigl( \begin{array}{l}
\psi_n \\
\psi_{n-1}
\end{array} \Bigr) ~.
\eeq
The allowed values of energy can therefore be found by computing the transfer 
matrix $M(q,E)$ obtained by multiplying together $P$ matrices \cite{ost},
\beq
M (q,E) ~=~ \prod_{n=1}^P ~\Bigl( \begin{array}{cc} 
-2E/J_n & - J_{n-1}/J_n \\
1 & 0 
\end{array} \Bigr) ~.
\label{trmat}
\eeq
Since ${\rm det} M(q,E) =1$ and ${\rm tr} M(q,E)$ is real, it is clear that
the eigenvalues of $M(q,E)$ must be either of the form $r$ and $1/r$, 
or of the form $\exp (ir)$ and $\exp (-ir)$, where $r$ is real
in either case. In the former case, the wave functions $\psi_n$ diverge
exponentially as $n \rightarrow \infty$ or $- \infty$; hence they are not
allowed physically. In the latter case, the wave functions do not diverge at 
infinity and are allowed. Thus, if $\vert {\rm tr} ~M(q,E) \vert = 2 \vert 
\cos r \vert \le 2$, the energy $E$ is allowed in the spectrum; otherwise that 
energy is not allowed. By following this method and sweeping through 
a large number of values of $q$ and $E$, we obtain the picture
of energy bands and gaps (shaded and unshaded regions, respectively) shown in 
Figs. 1 and 2, taking $\delta =0.05$. [We have sampled the energy values with 
a resolution of $dE = 10^{-6}$. It may be useful to clarify here that the 
numerous {\it vertical} gaps in Figs. 1, 2, 5, 6 and 7 have no physical 
significance; they are present because we have omitted the regions of 
$q/\pi$ whose rational 
approximations have such long periods $P$ that a numerical computation 
would have taken too long]. We have scaled both $E$ and $q$ in units 
of $\delta$ because the pictures turn out to depend only on those two 
ratios. We immediately see that the pictures look much more complicated
than the situation in which $q$ is {\it exactly} equal to zero; in that case,
there is precisely one gap extending from $\ed = 0$ to $1$. [We will consider 
only positive values of $q$ and $E$ since the spectrum is 
invariant under either $q \rightarrow -q$ (as is clear from (\ref{ham1}))
or $E \rightarrow -E$; the latter is clear from the invariance of 
$\vert {\rm tr} M(q,E) \vert$ under an unitary transformation by the Pauli
matrix
\beq
\sigma^3 ~=~ \Bigl( \begin{array}{cc}
1 & 0 \\
0 & -1
\end{array} \Bigr) 
\eeq
which reverses the sign and simultaneously converts $E \rightarrow - E$ 
for each of the $P$ matrices appearing in (\ref{trmat}). Thus the
energy bands shown in Figs. 1 and 2 should be understood as continuing to 
negative values of both $E$ and $q$ by reflection about those two axis].

Fig. 1 shows that as $\qd$ increases, the gaps shrink rapidly.
We will show that this can be understood using perturbation theory to $n^{\rm 
th}$ order, where $n$ is an odd integer, and that the gaps shrink as 
$(\delta /q)^n$. We have therefore labeled the gaps in Fig. 1 by the 
integers $n=1,3,5,...$. More interestingly, we observe that all the bands
approach the origin $(q,E)=(0,0)$. The widths of the low-lying bands vanish 
exponentially fast as $\qd \rightarrow 0$ as we will argue from a WKB 
analysis. This makes it impossible to find these portions of the bands by 
looking for energies satisfying $\vert {\rm tr} ~M(q,E) \vert \le 2$ using any 
reasonable energy resolution $dE$. We therefore find these thin portions of 
the bands by looking for {\it minima} of ${\rm tr} M^2 (q,E)$ as functions of 
$E$. If the energy resolution was infinitesimal, these minima would be found
at ${\rm tr} M^2 (q,E) = 2 \cos (2r) = -2$, i.e., ${\rm tr} M(q,E) = 2 \cos r
= 0$, which occurs inside an energy band. In practice, due to the finite 
energy resolution $dE$, these minima numerically yield single points in energy 
which lie within a distance of $dE$ of a band. We are thus able to find the 
locations of the thin regions without having to use a resolution smaller than
$dE = 10^{-6}$. In Fig. 2, we show these points (thin regions) which smoothly 
join on to the wider regions of the bands. We also show six curves which 
are the analytical results of a low-energy theory of the model which involves
solving a $1+1$-dimensional Dirac equation in a periodic potential. We will
also prove that the number of bands in the region $0 \le E/\delta \le 1$ 
increases as $2\delta /\pi q$ as $\qd \rightarrow 0$, while the number of 
states in each band is equal to $q/\pi$ when normalized appropriately in the 
thermodynamic limit, i.e., the number of sites $L \rightarrow \infty$. This 
will show that the number of states lying inside the $q=0$ gap is finite 
and equal to $2\delta /\pi^2$ in the limit $q \rightarrow 0$; this 
implies that $q=0$ is a rather singular point.

\subsection{Analytical results}

To begin, let us write the Hamiltonian in (\ref{ham1}) as a sum $H= 
H_0 + \vd$, where $\vd$ is the incommensurate term 
\beq
\vd ~=~ - ~\frac{\delta}{2} ~\sum_n ~\cos (Qn) ~(~ c_n^{\dagger} c_{n+1} ~+~ 
c_{n+1}^{\dagger} ~) ~,
\label{vterm}
\eeq
where $Q = \pi + q$ in this section. If 
\beq
\vert k \rangle ~=~ \frac{1}{\sqrt L} ~\exp (ikn) ~\vert n \rangle 
\eeq
denotes a momentum eigenstate of $H_0$, then the matrix elements of $\vd$ are 
given by
\beq
\langle k_1 \vert \vd \vert k_2 \rangle ~=~ - ~\frac{\delta}{4} ~(~ 
\delta_{Q+k_1,k_2} ~+~ \delta_{Q+k_2,k_1} ~) ~ (~ e^{-ik_1} ~+~ e^{ik_2} ~)~. 
\label{matel1}
\eeq

Let us understand the gaps for large values of $\qd$ by using 
perturbation theory about $\delta =0$. If $q$ and $\delta$ are both 
small, the states close to zero energy are dominated by momenta near $\pm 
\pi /2$. Since the incommensuration term $\vd$ in (\ref{vterm})
has Fourier components at momenta $\pm ( \pi + q)$, we see that the gaps above
zero energy result from the breaking of the energy degeneracy between the two 
states with momenta $k_1 = \pi /2 + nq/2$ and $k_{n+1} = -\pi /2 - nq/2$, 
where $n=1,3,5,...$. These two states are connected to each other through 
the $n-1$ successive intermediate states with momenta $k_2 = -\pi /2 + 
(n-2)q/2$, $k_3 = \pi /2 + (n-4)q/2$, $k_4 = -\pi /2 + (n-6)q/2$, ..., $k_n = 
\pi /2 - (n-2)q/2$, since we must have $k_{j+1} - k_j = \pm (\pi + q)$ mod 
$2\pi$. Since all the matrix elements $\langle k_{j+1} \vert \vd \vert k_j 
\rangle$ are approximately equal to $\pm i \delta /2$ (for $j=1,2,...,n$) for 
small values of $q$ in Eq. (\ref{matel1}), and all the energy denominators 
$E_{j} - E_1$ are of the order of $q$ (for $j=2,3,...,n$), we see that the 
two-fold degeneracy (at the energy $E_1 = E_{n+1} \simeq nq/2$) gets broken at 
the $n^{\rm th}$ order in perturbation theory to produce a gap of the order of 
$\delta^n /q^{n-1}$. To be explicit, we find that the upper and lower edges 
of the two gaps labeled by $n=1$ and $3$ in Fig. 1 are given by
\bea
\frac{E_{\pm} (1)}{\delta} ~&=&~ \frac{q}{2\delta} ~\pm ~ \frac{1}{2} ~, 
\nonumber \\
\frac{E_{\pm} (3)}{\delta} ~&=&~ \frac{3q}{2\delta} ~+~ \frac{3\delta}{16q} ~
\pm ~ \frac{\delta^2}{32q^2} ~, 
\eea
and we have verified that this agrees well with the numerically obtained 
boundaries. We can show that the general formula for the gap 
$\Delta E (n) = E_+ (n) - E_- (n)$ labeled by the odd integer $n=2p+1$ is 
given by
\beq
\frac{\Delta E (n)}{\delta} ~=~ \frac{2}{\delta} ~\vert ~\frac{\prod_{j=1}^n
\langle k_{j+1} \vert \vd \vert k_j \rangle}{\prod_{j=2}^n (E_1 - 
E_j )} ~\vert ~=~ \frac{( \delta /4q )^{2p}}{ ( p ! )^2} 
\eeq
to lowest order in $\delta$. 

We can count the number of states in the band lying between gaps $n$ and $n+2$ 
as follows. For $\delta =0$, let us normalize the number of states so that
the total number is $1$; since the momentum $k$ goes from $-\pi$ to $\pi$
in that case, the density of states in momentum space
is $1/(2\pi )$. If we now turn on a very small value of $\delta$, we find that
the band lying between the gaps $n$ and $n+2$ is made up of
linear combinations of the states lying between the two momenta intervals
$[\pi /2 + nq/2, \pi /2 + (n+2)q/2]$ and $[-\pi /2 - nq/2, -\pi /2 - 
(n+2)q/2]$. The total number of states in these two intervals is $2(q/2\pi) =
q/\pi$. Thus each band contains $q/\pi$ states. Now, this number cannot
change if we change $\delta$, and we can therefore use the same number
below in the opposite limit where $q/\delta$ is small.

Let us examine the more interesting regime where $\qd$ is small
and $\ed < 1$. For analyzing this, it is useful to consider the continuum 
limit. In this limit, the Fermi field $\psi (x) = c_n$ can be written as 
\beq
\psi (x) ~=~ \psr (x) ~\exp (\frac{i\pi x}{2}) ~+~ \psl (x) ~\exp ( - 
\frac{i\pi x}{2}) ~,
\label{fer1}
\eeq
where $\psr$ and $\psl$ denote the right- and left-moving fields, respectively;
they vary slowly on the scale of the lattice spacing. We substitute 
(\ref{fer1}) in Eq. (\ref{ham1}) and drop terms 
like $\exp (\pm i\pi x)$ which vary rapidly. We then find the following 
Dirac-like equations for the two time-dependent (Heisenberg) fields
\bea
i ~(~\partial /\partial t ~+~ \partial /\partial x ~) ~\psr ~-~ i \delta ~
\cos (qx) ~\psl ~&=&~ 0 ~, \nonumber \\
i ~(~\partial /\partial t ~-~ \partial /\partial x ~) ~\psl ~+~ i \delta ~
\cos (qx) ~\psr ~&=&~ 0 ~. 
\label{eom1}
\eea
The energy spectrum can be found from (\ref{eom1}) by defining 
$\psi_{\pm} = \psr \pm \psl$ which satisfy the equations
\bea
[ ~-~ \partial^2 /\partial x^2 ~+~ \delta^2 ~
\cos^2 (qx) ~-~ \delta q ~\sin (qx) ~ ] ~\psp ~&=&~ E^2 ~\psp~, \nonumber \\
{[} ~-~ \partial^2 /\partial x^2 ~+~ \delta^2 ~
\cos^2 (qx) ~+~ \delta q ~\sin (qx) ~ {]} ~\psm ~&=&~ E^2 ~\psm~. 
\label{eom2}
\eea
It is sufficient to solve one of these equations, say, for $\psp$, since 
$\psm$ is related to $\psp$ by
\beq
\psm (x) = \frac{i}{E} \Bigl[ ~- \partial /\partial x ~+~ \delta ~\cos (qx) ~
\Bigr] ~\psp (x) ~,
\label{pspm}
\eeq
provided $E \ne 0$. The energy spectrum can be found by solving the
time-independent equation
\beq
[~-~ \partial^2 /\partial x^2 ~+~ \delta^2 ~\cos^2 (qx) ~-~ \delta q ~\sin 
(qx) ~] ~\psp ~=~ E^2 ~\psp ~.
\label{eom3}
\eeq

Eq. (\ref{eom3}) has the form of a Schr\"odinger equation (with ``energy" 
$E^2$) in the presence of a periodic potential. The potential is similar but 
not identical to Mathieu's equation \cite{abr}. We therefore again expect 
bands to form. To simplify the notation, let us shift $x$ by $\pi /2q$ and 
then scale it by a factor of $q$ to make the period equal to $2\pi$. We then 
get
\beq
[~-~ \partial^2 /\partial x^2 ~+~ \frac{\delta^2}{q^2} ~\sin^2 x ~-~ \frac{
\delta}{q} ~\cos x ~] ~\psp ~=~ \frac{E^2}{q^2} ~\psp ~.
\label{eom4}
\eeq
By Floquet's theorem \cite{abr}, the solutions must satisfy $\psp (x+2\pi )=
e^{i\theta} \psp (x)$, where $\theta$ goes from $0$ to $\pm \pi$ from the
bottom of a band to the top. We observe that there is an exact zero energy 
state with $\psp (x) = \exp (~\frac{\delta}{q} \cos x~)$ and $\psm (x) =0$ for 
which $\theta =0$. In general, the nonzero energy states can only be found 
numerically. But if $\qd$ is large, the positions of the low-lying bands (with
$E^2 /q^2 \ll \delta /q$) can be found analytically as follows. For energies 
much lower than the maxima of the potential, we can ignore tunneling between 
the different wells to begin with. We note that there are two kinds of wells 
in Eq. (\ref{eom4}); those centered about $x=2m\pi$ and those centered about 
$x=(2m+1)\pi$, where $m$ can be any integer; the former wells are deeper than 
the latter. However, the {\it perturbative} energy levels (i.e., ignoring 
tunneling) are identical in the two types of wells, except for the $E=0$ state
which only exists in the deeper wells. This identity follows from 
Eqs. (\ref{eom2}) and (\ref{pspm}) which show that corresponding to any
(nonzero) energy eigenstate $\psp (x)$ which lies in a deeper well, there is 
an eigenstate $\psm (x)$ which lies in a shallower well. The $n^{\rm th}$ 
energy level in a deeper well has the same value as the $(n-1)^{\rm th}$ 
energy level in a shallower well, for $n \ge 1$. (This is an example of 
supersymmetric quantum mechanics \cite{wit}). It is therefore sufficient to
find the energy levels of (\ref{eom4}) which lie in one of the deeper wells, 
say, at $x=0$. Near the bottom of that well, we have a simple harmonic 
potential with small anharmonic corrections. Ignoring the anharmonic terms,
we find the energy levels to be simply given by
\beq
E_n^2 ~=~ 2 n q \delta ~,
\label{en1}
\eeq
where $n=0,1,2,...$. Next, we include the anharmonic corrections 
perturbatively. Up to second-order in perturbation theory, we get the more
accurate expression
\beq
E_n ~=~ \sqrt{2 n q \delta} ~\bigl[ ~1 ~-~ \frac{n}{8} ~\frac{q}{\delta} ~-~
\frac{5n^2 +2}{128} ~\frac{q^2}{\delta^2} ~\Bigr] ~.
\label{en2}
\eeq
In Fig. 2, we show the six curves corresponding to $n=1$ to $6$ in Eq. 
(\ref{en2}); for $n=0$, we simply get a straight line lying at zero energy. We 
see that all the curves agree extremely well with the numerical data, even up 
to $E_n$ of the order of $\delta$ where the harmonic approximation breaks down 
and the band widths become noticeable.

We will now consider tunneling between wells using the WKB method.
For $n \ll \delta /q$, the energy levels $E_n^2 /q^2$ of (\ref{eom4}) lie in 
small intervals centered about $x=m\pi$, where $m$ can be any integer
for $n \ge 1$ but must be an even integer for the $n=0$ band. In 
these small intervals, the potential is simple harmonic to a good 
approximation, with the harmonic frequency being equal to 
\beq
b \equiv \delta /q ~.
\eeq
We will assume that $b$ is large so that the band widths are small.
The splitting of the $n^{\rm th}$ band is given by the expression \cite{ter}
\bea
\frac{\Delta (E_n^2)}{q^2} ~&=&~ \frac{4b}{\pi} ~\exp (~- ~\int_{x_1}^{x_2} ~
dx ~{\sqrt {V(x) - 2nb}} ~)~, \nonumber \\
V(x) ~&=&~ b^2 ~\sin^2 x ~-~ b ~\cos x ~,
\label{wkb}
\eea
where $x_1$ and $x_2$ are the turning points, and we have approximated 
$E_n^2 /q^2$ by the lowest order result $2nb$ given in (\ref{en1}). For $n
\ge 1$, we take the turning points to be $x_1 \simeq \sqrt{(2n+1)/b}$ and $x_2
\simeq \pi - \sqrt{(2n-1)/b}$, while for $n=0$, $x_1 \simeq 1/\sqrt{b}$ and 
$x_2 \simeq 2\pi - 1/\sqrt{b}$.
To lowest approximation, the exponential factor in (\ref{wkb}) is 
given by $\exp (-2b)$ for $n \ge 1$ and by $\exp (-4b)$ for $n=0$. However,
since $\Delta (E_n^2) \simeq 2 E_n \Delta E_n$ for a small width
$\Delta E_n$ if $n \ge 1$, but $\Delta (E_0^2) = (\Delta E_0 )^2$ (since $E_0
=0$), we see that the splitting $\Delta E_n /q \sim \exp (-2b)$ for {\it all}
$n$. We can do a more accurate calculation to find the prefactors; we find 
that
\beq
\frac{\Delta E_n}{\delta} ~\sim ~ \frac{2^{3n+1}}{n ! \sqrt \pi} ~\Bigl(
\frac{\delta}{q} \Bigl)^{n-1/2} ~\exp (~ - \frac{2\delta}{q} ~) 
\label{wid1}
\eeq
for all values of $n$. We should note that the expression in (\ref{wid1})
for $n=0$ actually gives {\it half} the width of that band since the
$n=0$ band extends an equal amount to negative values of energy. 

Eq. (\ref{wid1}) explains why, with a numerical method using a finite energy 
resolution $dE$, the bands rapidly become thin and shrink to isolated points 
as $q \rightarrow 0$ in Fig. 2. In Fig. 3, we compare the numerically obtained 
widths of the lowest four bands ($n=0,1,2,3$) with the WKB expressions in 
(\ref{wid1}). The points in that figure indicate the thirty smallest values of 
width that we could find using the condition $\vert {\rm tr} ~M(q,E) \vert \le 
2$ with the energy resolution $dE = 10^{-6}$; the solid lines indicate the WKB 
expressions. We see that the agreement between the two becomes worse as $\qd$ 
increases, particularly for the larger values of $n$. This is expected because 
the WKB method is accurate only if $n \ll \delta /q$.

We will now prove that the number of states lying below the line $\ed =1$ 
is proportional to $\delta$. For large $\delta /q$, we can use a 
semiclassical phase space argument to count how many states lie below any
given energy. The ``Hamiltonian" on the left hand side of Eq. (\ref{eom3}) 
has the form $p^2 + V(x)$, where $V(x) = \delta^2 \cos^2 (qx) - \delta q 
\sin (qx)$. Hence the number of states up to energy $E$ is given by the 
phase space integral
\beq
\nu (E) ~=~ \frac{1}{L} ~\int \int \frac{dx dp}{2\pi} ~\Theta (~ E^2 - p^2 - 
V(x) ~) ~, 
\label{phsp}
\eeq
where we have divided by the length $L$ of the system for normalization;
the $\Theta$-function is defined to be $1$ and $0$ if its argument is
positive and negative, respectively. On doing the momentum integral in 
(\ref{phsp}), we get
\beq
\nu (E) ~=~ \frac{q}{2\pi^2} ~\int_0^{2\pi /q} ~dx ~\sqrt{E^2 - 
V(x)} ~\Theta (~E^2 - V(x) ~) ~,
\label{int1}
\eeq
where we have used the fact that $V(x)$ has period $2\pi /q$. For small values 
of $\qd$, we then see that the number of states lying between $E=0$ 
and $\delta$ is 
\beq
\nu (\delta ) ~=~ \frac{2\delta}{\pi^2} ~,
\label{num1}
\eeq
which is {\it independent} of $q$. Incidentally, this also yields the number 
of bands lying below $E=\delta$. Since each band contains $q/\pi$ states, 
Eq. (\ref{num1}) gives the number of bands to be $2\delta/\pi q$. We have 
verified that this estimate agrees very well with our numerical results.

We can obtain the density of states $\rho (E)$ from the expression in
(\ref{int1}) by differentiating it with respect to $E$. In the limit $\qd 
\rightarrow 0$, we get 
\bea
\rho (E) ~&=&~ ~\frac{2}{\pi^2} ~\int_0^U ~d\eta ~\Bigl[ ~1 ~-~
\Bigl(\frac{\delta}{E} \Bigr)^2 \sin^2 \eta \Bigr]^{-1/2} ~, \nonumber \\
{\rm where} \quad U ~&=&~ \sin^{-1} (\ed) \quad {\rm if} \quad E < \delta
\quad {\rm and} \quad \pi /2 \quad {\rm if} \quad E > \delta ~.
\label{den}
\eea
These are shown by the solid lines in Fig. 4; note the linear behavior
near $\ed =0$ and the logarithmic divergence
as $E \rightarrow \delta$ from either side. This is to be 
contrasted with the density of states found exactly at $q=0$. In that case, 
$\rho (E)$ vanishes for $E < \delta$ and is equal to $1/[\pi \sqrt{1 - 
(\delta /E)^2}]$ for $E > \delta$. This is shown by the dashed line in Fig. 4.

The difference in the spectrum for $q=0$ and $q \rightarrow 0$ should show
up most clearly in the specific heat at temperatures $T \ll \delta$. Since 
the chemical potential is zero, the free energy per site is given by
\beq
F ~=~ -2T ~\int_0^{\infty} ~dE ~\rho (E) ~{\rm ln} (~1 + e^{-E/T}~) ~,
\eeq
The factor of $2$ on the right hand side is because we have to sum over both
fermions and holes which have the same density of states. (We set the 
Boltzmann constant equal to $1$ for convenience). Then the specific heat per 
site is 
\beq
C_V ~=~ - ~T ~\frac{\partial^2 F}{\partial T^2} ~.
\eeq
For $q=0$, the
energy gap from $-\delta$ to $+\delta$ means that the specific heat vanishes
exponentially at low temperature. But for $q \rightarrow 0$, Eq. (\ref{den})
shows that the density of states goes linearly as $\rho (E) \simeq E/(\pi 
\delta)$ for $E \ll \delta$. This implies that the low-temperature specific 
heat goes as $T^2 /\delta$ which is very different from the exponential 
behavior at $q=0$.

We would like to point out here that the above results remain unchanged
if the incommensurate term is present in the chemical potential $\mu$ rather
than in the hopping $J_n$ in Eq. (\ref{ham1}), provided that $\delta$ and $q$ 
are both {\it small}. In other words, if we start with the Hamiltonian
\beq
H ~=~ - ~\frac{1}{2} ~\sum_n ~(~ c_n^\dagger ~c_{n+1} ~+~ c_{n+1}^\dagger 
c_n ~) ~-~ \delta ~\cos (\pi +q) n ~\sum_n ~c_n^\dagger ~c_n ~, 
\label{ham2}
\eeq
we get the same continuum theory as discussed above, if we define the Fermi 
fields as in Eq. (\ref{fer1}), followed by a phase redefinition $\psl 
\rightarrow i \psl$. This equivalence between the two continuum theories will
be used below.

\subsection{Reconciling $q=0$ with $q \rightarrow 0$: the Azbel-Hofstadter 
problem}

Let us ask: why does the model in Eq. (\ref{ham1}) show such 
different behaviors in the limit 
$q \rightarrow 0$ and at $q=0$? To answer this question, it is useful to
generalize the form of the incommensurate hopping in (\ref{ham1})
from $\delta \cos (\pi + q) n$ to $\delta \cos [(\pi + q)n +\eta ]$, where
$\eta$ is a phase lying between $0$ and $2\pi$. If $q/\pi$ is
{\it irrational} (this is the generic case to consider if $q$ is nonzero),
the spectrum does not depend on $\eta$. [If $q/\pi =M/N$ is rational, the 
spectrum does depend on $\eta$, but it varies less and less with $\eta$ as $N 
\rightarrow \infty$. In our numerical computations, we introduced a random 
phase $\eta$ for each value of $q$ and found that this had no noticeable 
effect on the two figures]. This insensitivity to $\eta$ can also be seen from 
the continuum theory in Eq. (\ref{eom1}) where a potential of the form 
$\cos (qx +\eta )$ can be transformed to $\cos (qx)$ by shifting $x$
appropriately. However, {\it exactly} at $q=0$, the hopping has a term like
$\delta \cos \eta \cos (\pi n)$ and the spectrum depends significantly on 
$\eta$; for instance, there is a gap from $E=0$ up to $\vert \delta \cos \eta 
\vert$, and the number of states up to an energy $E$ is given by
\beq
\nu (E, \eta ) ~=~ \frac{1}{\pi} ~\sqrt{E^2 ~-~ \delta^2 \cos^2 \eta} ~
\label{num2}
\eeq
for $\vert \delta \cos \eta \vert < E \ll 1$. We may now consider taking an 
{\it average} of this number over all possible values of $\eta$. Thus, the 
number of states up energy $E$ is given by Eq. (\ref{num2}) to be
\beq
\nu (E) ~=~ \int_0^{2\pi} ~\frac{d\eta}{2\pi^2} ~\sqrt{E^2 - \delta^2 \cos^2 
\eta} ~\Theta (~ E^2 - \delta^2 \cos^2 \eta ~)~.
\eeq
This agrees with (\ref{int1}) if we substitute $\eta =qx$. Thus, 
the limit $q \rightarrow 0$ agrees with the point $q=0$ {\it provided} that 
we average over $\eta$ in the latter case. 

The question now arises: is there a physical system where an average over the 
phase $\eta$ occurs naturally? Some experimental systems sitting at $q=0$ (the 
dimerized point) are more likely to choose a particular value of $\delta$ 
with $\eta =0$, rather than average over many values 
of $\eta$; the limit $q \rightarrow 0$ and $q$ exactly equal to zero will
therefore differ from each other. For simplicity, we will choose $\eta =0$ 
in the following sections to describe such systems.

There is, however, a problem in {\it two} dimensions which averages over 
$\eta$ in a natural way. This is the Azbel-Hofstadter (AH) problem in which
electrons hop on a regular lattice in the presence of an uniform magnetic 
field pointing normal to the plane. Let us consider a square lattice with the
sites being labeled by two integers $(m,n)$. (We set the lattice spacing
equal to $1$ as usual). The hopping element from the site $(m,n)$ to the site
$(m+1,n)$ will be denoted by $t(m,n;{\hat x})$, while hopping from 
$(m,n)$ to $(m,n+1)$ will be denoted by $t(m,n;{\hat y})$; these will 
be complex numbers in general. Suppose that the hopping elements are all real
in the absence of a magnetic field, with $t(m,n;{\hat x}) =t_1$ and $t(m,n;
{\hat y}) =t_2$. In the presence of a magnetic field, let $\Phi$ denote the
magnetic flux through each square. Then some of the hopping elements must
necessarily be taken to be complex, with the sum of the phases of the four
hopping elements taken anticlockwise around each square being equal to 
$Q = e\Phi/(\hbar c)$, where $e$ is the electronic charge and $c$ is the 
speed of light. We can interpret $Q /(2\pi)$ as the ratio of the flux 
through each square to the flux quantum $\Phi_0 = 2 \pi \hbar c/e$.
Clearly, $Q$ and $Q + 2 \pi$ are physically 
indistinguishable; thus, $Q = \pi$ corresponds to the largest magnetic 
field that can appear on a lattice. Now, there are many choices possible for 
distributing the phases amongst the different hopping elements; the 
different choices are related to each other by gauge transformations. Let us 
choose the Landau gauge in which the hopping elements in the $\hat y$ 
direction stay real and are equal to $t_2$, while the hopping elements in the
$\hat x$ direction are given by
\beq
t(m,n;{\hat x}) ~=~ t_1 ~\exp (-i Q n) ~.
\eeq
The lattice Schr\"odinger equation satisfied by the single-particle wave
functions is then given by
\beq
t_1 ~(~ \exp (iQ n) ~\phi_{m+1,n} ~+~ \exp (-iQ n) ~\phi_{m-1,n} ~
)~+~ t_2 ~(~ \phi_{m,n+1} ~+~ \phi_{m,n-1} ~) ~=~ e ~\phi_{m,n} ~.
\label{hofs}
\eeq
Since the hopping elements do not depend on the $\hat x$-coordinate $m$, we 
may assume the wave functions to have the factorized form 
\beq
\phi_{m,n} ~=~ \exp (i m \eta ) ~\psi_n ~,
\eeq
where the momentum $\eta$ can take any value from $0$ to $2\pi$. Eq. 
(\ref{hofs}) then takes the form 
\beq
- \frac{1}{2} ~(~ \psi_{n+1} ~+~ \psi_{n-1} ~)~ -~ \frac{t_1}{t_2} ~\cos 
(Q n ~+~ \eta )~ \psi_n ~=~ - \frac{e}{2t_2} ~\psi_n ~.
\label{harp}
\eeq
This is the well-studied Harper's equation with an on-site incommensurate term. 
Further, we see that a phase $\eta$ appears naturally in that term, and that 
we have to consider all values of $\eta$ lying in the interval $[0,2\pi ]$. 

Let us consider the extremely anisotropic limit $\delta \equiv t_1 /
t_2 << 1$. We will show that this problem has the same continuum limit as 
the problem in which the incommensurate term is in the hopping. If we define
\beq
\delta ~=~ \frac{t_1}{t_2} \quad {\rm and} \quad E ~=~ - ~\frac{e}{2t_2} ~,
\eeq
then the Hamiltonian corresponding to Eq. (\ref{harp}) can be written as
$H=H_0 + \vd$, where the incommensurate term is
\beq
\vd ~=~ - ~\delta ~\sum_n ~\cos (Qn + \eta) ~c_n^{\dagger} c_n ~.
\eeq
The matrix elements of $\vd$ between momentum eigenstates of $H_0$ are given
by
\beq
\langle k_1 \vert \vd \vert k_2 \rangle ~=~ - ~\frac{\delta}{2} ~(~ 
\delta_{Q+k_1,k_2} ~e^{-i\eta} ~+~ \delta_{Q+k_2,k_1} ~e^{i\eta} ~)~. 
\label{matel2}
\eeq
For $Q$ close to $\pi$ and $k_1 , k_2$ close to $\pm \pi /2$, this has,
up to factors of $i$, the
same form as (\ref{matel1}), except that we have to average over $\eta$ in the
AH problem. We thus see that the anisotropic limit of the AH problem provides 
us with a physical realization of Eq. (\ref{ham1}) except that we have to 
consider a phase average of the incommensurate term. Dimerization
in the one-dimensional problem corresponds to $Q = \pi$, namely, 
the magnetic flux through each square of the two-dimensional lattice is equal
to half a flux quantum. We may therefore carry over all our results to that 
particular limit of the AH problem. For instance, the density of states for 
$Q \rightarrow \pi$ has the form given by the solid lines in Fig. 4. Note 
that the energy spectrum does not depend on the phase $\eta$ for generic 
(i.e., irrational) values of $Q$ in Eq. (\ref{harp}). So if $L_x$ 
denotes the number of lattice sites in the $\hat x$-direction, the number 
of values of $\eta$ is also equal to $L_x$ since $\eta$ takes all 
values from $0$ to $2\pi$ in steps of $2\pi /L_x$.
Hence the density of states for the AH problem is equal to $L_x$ times the
density given in Fig. 4 for the one-dimensional problem.

We note that the energy eigenvalues in Eq. (\ref{hofs}) are invariant
under the exchange $t_1 \leftrightarrow t_2$ corresponding to an interchange
of the $\hat x$ and $\hat y$ hopping elements in the Landau gauge. Thus the 
AH problem has the duality symmetry $\delta \rightarrow 1/\delta$. Our 
problem with an incommensurate hopping has no such symmetry.

With the present day expertise in fabricating quantum dot arrays, it may be
possible to construct a two-dimensional lattice whose spacing is large enough 
that the flux through each square could be made equal to half a flux quantum 
with currently available magnetic field strengths. (A similar suggestion was 
made by Hofstadter \cite{azb}). Consequently, our expression for the density
of states of such a system could be experimentally tested.

\section{Interacting fermions close to dimerization}

Let us now return to one dimension. In this section,
we will argue that the difference between small $q$ and $q=0$ (in
particular, the presence of states within the $q=0$ gap) persists for the
more interesting case of interacting fermions, i.e., for Luttinger liquids.
To this end, consider adding a four-fermion interaction
term like $\sum_n c_n^\dagger c_n c_{n+1}^\dagger c_{n+1}$ to the
Hamiltonian (\ref{ham1}). Equivalently, we can consider an $XXZ$ spin-$1/2$
chain governed by
\beq
H ~=~ \sum_n ~\Bigl[ ~( ~1 ~+~ \delta \cos (\pi + q)n ~) ~
(~S_n^x S_{n+1}^x ~+~ S_n^y S_{n+1}^y ~) ~+~ D ~S_n^z S_{n+1}^z ~
\Bigr] ~.
\label{ham3}
\eeq
which is related to the interacting fermion theory by a Jordan-Wigner
transformation \cite{lie}. If the incommensurate term is absent ($\delta =0$), 
it is known from the exact Bethe ansatz solution and conformal field theory
\cite{sch} that the model in (\ref{ham3}) is gapless 
with a linear dispersion of the low-energy excitations if 
$-1 < D \le 1$. If we now add a dimerization ($q=0$ and $\delta$ is
small), then the system becomes gapped and the gap scales as
\bea
\Delta E ~&=&~ \delta^{1/(2-K)} ~, \nonumber \\
{\rm where} \quad K ~&=&~ \frac{\pi}{2 \cos^{-1} (-D)} ~,
\label{gap}
\eea
where $0 < \cos^{-1} (-D) \le \pi$.
[Actually, a gap opens up only if $K \le 2$. If $K > 2$, the dimerization term
is irrelevant in the sense of the renormalization group, and a gap is not
generated]. It is useful to state the result in Eq. (\ref{gap}) in the 
language of bosonization. We introduce a bosonic field $\phi (x,t)$ such
that bilinears in Fermi fields have local expressions in terms of the
bosonic field \cite{sch}. For instance,
\bea
\psi_R^\dagger \psl ~&\sim &~ \exp (i2\phi ) ~, \nonumber \\
\psi_L^\dagger \psr ~&\sim &~ \exp (-i2\phi ) ~,
\label{bos1}
\eea
and
\beq
\partial \phi /\partial x ~=~ - ~\pi ~[ ~\rho (x) ~-~ \rho_0 ~] ~,
\label{bos2}
\eeq
where $\rho (x)$ is the local fermion density, and $\rho_0$ is the average
density. For $q=0$, the model in (\ref{ham3}) is equivalent, at low energies, 
to a sine-Gordon theory described by the Lagrangian density \cite{sch}
\beq
{\cal L} ~=~ \frac{1}{2\pi vK} ~\Bigl[ ~(\partial \phi /\partial t)^2 ~-~ 
v^2 ~ (\partial \phi /\partial x)^2  ~\Bigr] ~-~ \alpha ~\delta^{2/(2-K)} ~
[~1 ~-~\cos (2 \phi) ~] ~,
\label{lag1}
\eeq
where $v$ is the velocity of low-energy excitations and
$\alpha$ is a positive constant; their numerical values are not important
here. The main point is to note the exponent of $\delta$ in the coefficient
of $\cos (2\phi)$. (There are also factors of $\ln \delta$ due to the presence
of a marginal operator in the $XXZ$ model \cite{aff}, but we will ignore such 
terms here). The number $K$ and the velocity $v$ are the two important
parameters which determine the low-energy, long wavelength behavior
of a Luttinger liquid.

It is important to observe that the incommensurate term which is {\it linear} 
in $\delta$ in the original fermionic theory in (\ref{ham3}) becomes a cosine
term with a different exponent for $\delta$ in the low-energy bosonic theory. 
This is because a nontrivial renormalization occurs in the process of deriving 
the low-energy bosonic theory from the microscopic fermionic theory 
\cite{bha,wie}. This
renormalization occurs even if the fermions are noninteracting, i.e.,
for the $XY$ spin-$1/2$ chain with $D =0$ in (\ref{ham3}) and $K=1$; the
reason for this is that the sine-Gordon theory is always strongly interacting. 
These strong interactions are also responsible for the large renormalizations 
of the correct quantum spectrum compared to the naive (i.e. classical)
spectrum that one obtains from the sine-Gordon theory, 
namely, the classical soliton mass for the fermionic excitations and the 
quadratic fluctuations around $\phi =0$ spectrum for the bosonic excitations
\cite{raj}. The noninteracting fermionic model in (\ref{ham1}) has no such 
renormalizations, which is why we did not use bosonization in the earlier
part of this paper.

Let us now change $q$ slightly away from zero. Since $\cos (\pi + q)n = \cos 
(\pi n) \cos (q n)$ on a lattice, we have a dimerization whose coefficient 
$\delta \cos qn$ varies very slowly over the system. Thus, the low-energy 
excitations, which had a relativistic dispersion with the mass
$\delta^{1/(2-K)}$ for $q=0$, now have a space-dependent mass $\vert \delta
\cos (qx) \vert^{1/(2-K)}$. To put it differently, bosonization 
yields a theory of the sine-Gordon type, except that the coefficient of 
the $\cos (2\phi )$ term in Eq. (\ref{lag1}) gets modified to produce
\beq
{\cal L} ~=~ \frac{1}{2\pi vK} ~\Bigl[ ~(\partial \phi /\partial t)^2 ~-~
v^2 ~ (\partial \phi /\partial x)^2  ~\Bigr] ~-~ \alpha ~\vert \delta \cos 
(qx) \vert^{2/(2-K)} ~[~1 ~-~ \cos (2 \phi) ~] ~,
\label{lag2}
\eeq
Unlike (\ref{lag1}), the theory in Eq. (\ref{lag2}) cannot be solved 
analytically, 
either in classical or in quantum mechanics. However we can make some 
qualitative statements about the low-energy spectrum if $\qd \ll 1$. When
calculating the spectrum of small oscillations about a classical ground state
$\phi =0$, we find it convenient to first shift $x$ by $\pi /(2q)$ to 
change $\cos (qx)$ to $\sin (qx)$, and to then scale $x$ by a factor of
\beq
a ~=~ (q \delta )^{1/(3-K)} ~.
\eeq
This gives the eigenvalue equation
\beq
-~v^2 ~\frac{\partial^2 \phi}{\partial x^2} ~+~ 4 \pi K \alpha v ~\vert x 
\vert^{2/(2-K)} ~ \phi ~=~ \Bigl( \frac{E}{a} \Bigr)^2 ~\phi 
\label{schr}
\eeq
for eigenstates lying close to the origin $x=0$; we have approximated $\sin
(qx)$ by $qx$ and $\sin (2\phi )$ by $2\phi$. Since (\ref{schr}) is the
Schr\"odinger equation with a confining potential, we see that the energy 
can take several discrete values which are given by numerical factors 
multiplying $a$. [Note that our earlier results for noninteracting fermions 
agree with this scaling argument if we set $K=1$]. These discrete values will 
then spread out into bands once we include tunneling between the different 
wells centered at the points $x=m \pi$. Similarly, we can 
find the energy of a classical soliton which goes from $\phi =0$ at $x 
\rightarrow - \infty$ to $\phi = \pi$ at $x \rightarrow \infty$; once again 
we can show by scaling that this will be given by $a$ times 
some numerical factor. [Of course, all the numerical factors will get
renormalized due to quantum corrections, but the power-law dependence on 
$q\delta$ is not expected to change]. We therefore see that all the low-lying 
excitations have energies of the order of $(q \delta)^{1/(3-K)}$; this is 
much smaller than the gap which exists exactly at $q=0$ given by 
Eq. (\ref{gap}), since we are assuming that $\qd$ is small. We thus see that 
in the limit $q \rightarrow 0$, there are many states which lie within that
gap. Indeed, we can use WKB quantization for Eq. (\ref{schr})
to estimate that the number of bands lying between zero energy and the 
gap $\Delta E$ in (\ref{gap}) is of the order of $\Delta E /q$, in the limit
$q \rightarrow 0$. The number of states in the same interval is therefore of 
the order of $\Delta E$; this also follows from the semiclassical argument
presented below.

The semiclassical phase space estimate given in Eq. (\ref{int1}) can be used
to find the density of states $\rho (E)$ at energies much smaller than the 
$q=0$ gap. If $E \ll \delta^{1/(2-K)}$, the wave function for that state 
lies in a region where we can approximate $\sin (qx)$ by $qx$. Ignoring 
various numerical factors, we then get
\beq
\rho (E) ~=~ \frac{d}{dE} ~\int_0^{2\pi /q} ~\frac{q dx}{\pi} ~\sqrt{E^2 - 
(\delta q x)^{2/(2-K)}} ~\Theta (~E^2 - (\delta q x)^{2/(2-K)} ~) ~\sim ~
\frac{E^{2-K}}{\delta} ~.
\label{int2}
\eeq
Since the chemical potential for these bosonic excitations is zero, the free
energy per site is given by
\beq
F ~=~ T ~\int_0^{\infty} ~dE ~\rho (E) ~{\rm ln} (~1 - e^{-E/T} ~)~.
\eeq
The specific heat at temperatures much lower than the $q=0$ gap therefore
scales as $T^{3-K} /\delta$, which is again much larger than the exponential 
dependence which occurs at $q=0$. [It is tempting to compare this exponential 
versus power-law dependence of the specific heat with the forms of the 
specific heats of the magnetic excitations observed in the dimerized ($q=0$) 
and incommensurate ($q \ne 0$) phases of $CuGeO_3$ \cite{lor}. However
that system is much more complicated because it has strong phonon-spin 
couplings; we have ignored such terms in our model by assuming that the 
incommensuration is static].

We should note that Eq. (\ref{int1}) counts the number of one-particle 
fermionic states, and the specific heat calculated from that gets a 
contribution from states with all possible fermion numbers. Eq. (\ref{int2}),
however, only counts the number of one-particle bosonic states, i.e., 
fermion-hole excitations. Hence the specific heat calculated from that only 
gets a contribution from states with zero fermion number. However the 
complete expression for the specific heat is expected to differ only by 
numerical factors from the bosonic one, and should therefore have the same 
kind of exponential or power-law dependence on the temperature.

\section{Fermions near other commensurate fillings}

It is interesting to note that the peculiar difference 
between the limit $q \rightarrow 0$ and $q=0$ only occurs near dimerization,
i.e., near wave number $\pi$. There is no such singularity at $q=0$ if we take 
the incommensurate term in Eq. (\ref{ham1}) to be of the form $\delta \cos 
(Qn)$, where $Q=Q_0 +q$ and $Q_0 /\pi$ is equal to some simple rational number 
{\it different} from $1$. Let us first see this for the case of 
noninteracting fermions. Imagine filling up the Fermi sea up to a Fermi 
energy $E_F = -\cos k_F$ where $k_F = Q_0 /2$; the Fermi velocity is then $v 
= \sin k_F$. We define the continuum Dirac field as 
\beq
\psi (x) ~=~ \psr (x) ~\exp ( i k_F x) ~+~ \psl (x) ~\exp ( - i k_F x) ~,
\label{fer2}
\eeq
After dropping terms like $\exp (\pm i k_F x)$ which vary rapidly, the Dirac 
equations take the form
\bea
i ~(~\partial /\partial t ~+~ v \partial /\partial x ~) ~\psr ~+~ 
\frac{\delta}{2} ~\exp (iqx-ik_F) ~\psl ~&=&~ 0 ~, \nonumber \\
i ~(~\partial /\partial t ~-~ v \partial /\partial x ~) ~\psl ~+~ 
\frac{\delta}{ 2} ~\exp (-iqx+ik_F) ~\psr ~&=&~ 0 ~. 
\label{eom5}
\eea
The solutions of this equation have the plane wave form
\beq
\Bigl( \begin{array}{c} 
\psr (x,t) \\
\psl (x,t)
\end{array} \Bigr) ~=~ \exp (-iEt) ~\Bigl( \begin{array}{c} 
a_R ~\exp (ikx + iqx/2 - ik_F/2) \\
a_L ~\exp (ikx - iqx/2 + ik_F/2) 
\end{array} \Bigr) ~.
\label{trans}
\eeq
On substituting this in (\ref{eom5}), we find that there is a single energy 
gap lying between $E_+$ and $E_-$ given by
\beq
E_{\pm} ~=~ \frac{1}{2} ~(~v q ~\pm ~ \delta ~) ~.
\label{epm}
\eeq
Thus the size of the gap, $\delta$, is independent of $q$. (All these 
calculations 
assume that $q$ and $\delta$ are both small; the gap size may depend on $q$ if 
$q$ is not small). If $\delta$ is held fixed and $q$ is increased from zero, 
a state first appears at zero energy when $q$ reaches the value $\delta /v$. 
It is instructive to express all this in the language of bosonization. Since
$K=1$, we can combine Eqs. (\ref{bos1}) and (\ref{eom5}) to show that the 
sine-Gordon Lagrangian takes the form
\beq
{\cal L} ~=~ \frac{1}{2\pi v} ~\Bigl[~ (\partial \phi /\partial t)^2 ~-~ 
v^2 ~(\partial \phi /\partial x)^2  ~\Bigr] ~-~ \frac{\pi 
\delta^2}{16 v} ~[~1 ~-~ \cos (2 \phi + qx - k_F ) ~] ~.
\eeq
Note that we have fixed the coefficient of $\cos (2\phi )$ in such a way that,
for $q=0$, the soliton mass, including the quantum corrections \cite{raj}, 
is exactly equal to the gap from zero energy, i.e., $\delta /2$. On shifting
$\phi$ by $(qx- k_F )/2$, we get
\beq
{\cal L} ~=~ \frac{1}{2\pi v} ~\Bigl[ ~(\partial \phi /\partial t)^2 ~-~ 
v^2 ~(\partial \phi /\partial x ~-~ q/2 )^2  ~\Bigr] ~-~ \frac{\pi 
\delta^2}{16 v} ~[~1 ~-~ \cos (2 \phi ) ~] ~.
\eeq
The Hamiltonian therefore contains a boundary term 
\beq
H_b ~=~ - ~\frac{qv}{2\pi} ~[~\phi (\infty ) ~-~ \phi (-\infty ) ~]
\label{bound}
\eeq
which is purely topological; it is equal to $-qv /2$ in a 
one-soliton state. We thus see that if $q$ is increased from zero, the 
energy of the one-soliton state becomes equal to that of the ground
state (whose soliton number is zero) when $q$ reaches the value $\delta /v$.
From (\ref{bos2}), we observe that
a soliton corresponds to a hole; thus the energy of a hole (i.e., 
the energy required to remove one fermion from the system) becomes zero at 
$q=\delta /v$. Conversely, the energy required to add a fermion 
(antisoliton) to the system becomes zero at $q=-\delta /v$. [Similar 
results appear in the context of incommensurate crystals \cite{lyu,bha,nak}]. 

To illustrate these results, we show the two major bands and the big
gap separating them near the wave number $Q_0 = 2\pi /3$ and the 
Fermi energy $E_F = - \cos (\pi /3) = -1/2$ in Fig. 5;
we have taken $\delta =0.4$. [We have chosen a bigger value of $\delta$
here compared to Figs. 1 and 2 in order to discuss some additional gaps
in Sec. V and Fig. 6 below; those small gaps would not have been visible in
Fig. 5 if we had chosen values of $\delta$ much less than 
$0.4$]. We used the condition ${\rm tr} ~M(q,E) \le 2$
to find the energy levels, and we used the energy resolution $dE = 10^{-5}$. 
Note that the upper and lower edges of the gap follow Eq. (\ref{epm}), and
the gap closes with respect to $E_F$ at $\qd = \pm 1/v$, where $1/v = 
2/{\sqrt 3} = 1.1547...$. It is remarkable how simple Fig. 5 is compared to 
Figs. 1 and 2.

Thus, if $Q_0 \ne \pi$, there is a critical and nonzero value of $q$ at which 
the gap closes at zero energy. This is very
different from the earlier case with $Q_0 =\pi$, where there is always 
a band of energies lying around zero energy, so that there is no
gap at zero energy for any $q \ne 0$. (Our results disagree with Ref. 
8 which argues that there is a nonzero critical value of $q$ for 
$Q_0 =\pi$).

Finally, we can show that there is a nonzero critical value of $q$ where the
gap closes if $Q \ne \pi$ and the fermions are {\it interacting}. By following
arguments similar to the ones above, we find that the continuum theory 
is described by a combination of the Lagrangian in (\ref{lag1}) (for $q=0$) 
and the boundary term in (\ref{bound}) (for $q$ nonzero but small). The
soliton gap for $q=0$ is given by $\Delta E \sim \delta^{1/(2-K)}$; hence the 
boundary term implies that the soliton (antisoliton) gap will become zero 
at $q = \pm q_c$ where
\beq 
q_c ~=~ \frac{2K\Delta E}{v} ~.
\eeq
We thus get an expression for the critical value of $q$ in terms of the 
$q=0$ gap and the two Luttinger parameters $K$ and $v$.

\section{Higher order continuum theories for noninteracting fermions with
incommensurate hopping}

In this section, we will consider the complete energy spectrum of 
noninteracting fermions with a weak incommensurate hopping. (This will 
complement the pictures of the complete spectrum of noninteracting fermions 
with an incommensurate on-site potential given in the works of Hofstadter and 
of Sun and Ralston \cite{azb}). In contrast to Secs. II and IV, we will use
continuum theories which go up to higher than linear order in $\delta$. 
This will enable us to look at some of the smaller gaps in the energy 
spectrum as can be seen in Fig. 7.

The basic point is the following. Suppose that the incommensurate term $\vd$ 
has Fourier components at wave numbers $\pm Q$ (where $Q= Q_0 +q$), and 
that we are interested in the vicinity of the two 
Fermi points at momenta $\pm k_F$. (We assume that $0 < Q_0 < 2\pi$ and
$0 < k_F < \pi$). Then a gap develops at the Fermi 
points if there exists an integer $m$ (positive or negative) such that
\beq
- ~k_F ~+~ m Q_0 ~=~ k_F \quad {\rm mod} \quad 2 \pi ~.
\label{cond1}
\eeq
To lowest order in the strength of the incommensuration $\delta$, the gap
at the Fermi points is of order $\delta^{\vert m \vert}$, since 
$\vd$ must act $\vert m \vert$ times to take us from the momentum 
state $-k_F$ to $k_F$ or vice versa. Now suppose that there is another
integer $n$ such that
\beq
- ~k_F ~+~ n Q_0 ~=~ k_F \quad {\rm mod} \quad 2 \pi ~.
\label{cond2}
\eeq
We now have the possibility of
interference between the actions of $\vd^{\vert m \vert}$ and $\vd^{\vert n 
\vert}$, i.e. between terms in the continuum theory of the forms
$\delta^{\vert m \vert} e^{imqx}$ and $\delta^{\vert n \vert} e^{inqx}$. 
This interference gives rise to a large number of gaps and bands as we 
can show through a harmonic potential approximation. It is clear that 
conditions (\ref{cond1}) and (\ref{cond2}) can hold simultaneously only 
if $Q_0 /\pi$ and $k_F /\pi$ are rational numbers.

There are now two possibilities which we will call A and B. In case A, 
$m = -n$, which necessarily means that $k_F = \pi /2$ and $E_F =0$. In that 
case, all the gaps are of the order $\delta^{\vert m \vert}$. 
The discussion in Sec. II B and Figs. 1 and 2 is an example of case 
A, with $Q_0 = \pi$, $k_F = \pi /2$ and $m=-n=1$. In case B, $\vert m \vert 
< \vert n \vert$. Here, there is one big gap of order $\delta^{\vert m 
\vert}$ and all the other gaps are of order $\delta^{\vert n \vert}$.

As an example of case B, let us consider the 
two small gaps originating from the top and bottom of the big gap shown in 
Fig. 5 at $\qd =0$. We will only discuss the gap at the top here. An 
enlargement of that region is shown in Fig. 6 where it is clear that there 
are actually several small bands and gaps emerging from 
the top of the big gap; only the largest of these small gaps is visible in 
Fig. 5. This region can be understood as follows. Following the notation in
Eqs. (\ref{cond1}-\ref{cond2}), we have $Q_0 = 2\pi /3$, $k_F = \pi /3$, 
$m=1$ and $n=-2$. As discussed in Sec. IV, the effect of (\ref{cond1}) 
is linear in $\delta$. We now have to consider the effect of (\ref{cond2})
which is of second order in $\delta$. The fact that two applications of $\vd$
can take us from the modes near $-k_F$ to the modes near $k_F$ has three 
different effects.

\noindent (i) At $q=0$, it shifts the position of the lower edge of the 
upper band from the value $E_{q=0} = E_F + E_+ = -1/2 + \delta /2$ given 
in Eq. (\ref{epm}) with $E_F = -1/2$. This shift can 
actually be calculated exactly, not just to order $\delta^2$, by considering 
the three states with momenta $-\pi /3$, $\pi /3$ and $\pi$ and writing down
the matrix elements of $H_0 + \vd$ between these states as a $3 \times 3$ 
matrix. One of the eigenvalues of this matrix gives the position of the
lower edge of the upper band to be
\beq
E_{q=0} ~=~ \frac{1}{4} ~+~ \frac{\delta}{4} ~-~ \frac{1}{4} ~\Bigl( ~9 ~-~ 6
\delta ~+~ 3\delta^2 ~\Bigr)^{1/2} ~.
\label{leub}
\eeq

\noindent (ii) At second-order in $\vd$, we can go from the a state near
$-\pi /3$ to a state near $\pi$ and then back to the original state near
$-\pi /3$; similarly, we can go from $\pi /3$ to $\pi$ and back to $\pi /3$.
This changes the velocity $v = \partial E /\partial k$ from $v = {\sqrt 3}/2$
to
\beq
v ~=~ \frac{\sqrt 3}{2} ~\Bigl( ~1 ~-~ \frac{\delta^2}{36} ~\Bigr) ~.
\label{vel}
\eeq

\noindent (iii) Finally, we can define the continuum Dirac fields as in 
(\ref{fer2}). After performing the transformation in (\ref{trans}) and 
shifting $x$ by an appropriate amount, we get the time-independent equations
\bea
[~ {\tilde E} ~+~ i v \partial /\partial x ~]~ a_R ~+~ \frac{\delta}{2} ~[~
1 ~-~ \frac{\delta}{12} ~\exp (-i3qx) ~]~ a_L ~&=&~ 0 ~, \nonumber \\
{[}~ {\tilde E} ~-~ i v \partial /\partial x ~]~ a_L ~+~ \frac{\delta}{2} ~[~
1 ~-~ \frac{\delta}{12} ~\exp (i3qx) ~]~ a_R ~&=&~ 0 ~, 
\label{eom6}
\eea
where
\bea
{\tilde E} ~&=&~ E ~-~ \frac{vq}{2} ~-~ E_{q=0} ~+~ \beta ~, \nonumber \\
\beta ~&=&~ \frac{\delta}{2} ~(~ 1 ~-~ \frac{\delta}{12} ~)~.
\eea
Thus we get a periodic potential as in Sec. II B. Following the procedure in
that section, we first study the harmonic oscillator energies near the
bottom of one of the wells, say, at $x=0$; then we will briefly comment on the
broadening of those energies due to tunneling between different wells.
On expanding $\exp (\pm i3qx)$ up to order $q^2$, we find that the fields 
$a_{\pm} = a_R \pm a_L$ satisfy the harmonic oscillator equations
\bea
[~ - ~\frac{1}{2} ~\partial^2 /\partial x^2 ~+~ \frac{1}{2} ~\omega^2 x^2 ~
+~ \frac{\beta^2 - {\tilde E}^2}{2v^2} ~]~ a_{\pm} ~=~ 0 ~,
\eea
where the harmonic frequency is
\beq
\omega ~=~ \frac{\sqrt 3}{4} ~\frac{\delta^{3/2} \vert q \vert}{v} ~.
\eeq
The harmonic approximation is good if the root mean square deviation of the 
particle $\langle x^2 \rangle^{1/2} \sim 1/{\sqrt \omega}$ is much smaller 
than $2\pi /3q$ which is the periodicity of the potential in (\ref{eom6}). We 
therefore assume that
\beq
v ~\vert q \vert \ll \delta^{3/2} ~.
\eeq
We then find the energy levels to be
\beq
E_n ~=~ E_{q=0} ~+~ \frac{vq}{2} ~-~ \beta ~+~ \Bigl[ ~\beta^2 + 2v^2 \omega 
(n+ \frac{1}{2} )~ +~ \frac{vq\delta^2}{8} ~\Bigl]^{1/2} ~,
\label{en3}
\eeq
where $n \ge 0$. The term $vq\delta^2 /8$ inside the square root comes 
from perturbatively including the first anharmonic correction. The 
eight solid lines in Fig. 6 show the energies in (\ref{en3})
for $n=0,1,2,3$, and for positive and negative values of $q$.
The agreement is good for small values of $q$. We can now use the WKB
approximation to study the band widths due to tunneling between wells; we
find that the widths go as 
\beq
\Delta E_n ~\sim ~ \exp (~-\frac{4}{9} ~\frac{\delta^{3/2}}{\vert q \vert}~)~.
\label{wid2}
\eeq
If $\delta$ and $\qd$ are both small, then (\ref{wid2}) is much larger than 
the factor of $\exp (-2 \delta /q)$ obtained in Eq. (\ref{wid1}).
Qualitatively, this explains why the bands in Fig. 6 broaden much more 
rapidly than the ones in Fig. 2 as $q$ increases from zero.

Fig. 7 shows the complete picture of bands and gaps for all values
of $Q$ with $\delta = 0.4$. We see that the gaps come in two classes. As 
discussed above, a gap
arising from case B consists of a small gap (which, at higher resolution, is 
actually a group of bands and gaps) emerging from one point on one edge of 
a big gap. On the other hand, a gap arising from case A looks like 
a cross in which four groups of gaps meet at a point; this can 
be seen in Fig. 7 at $E=0$ and $Q/\pi = 1/m$ where $m=1,2,3,...$. The simplest
example of this corresponds to $m=1$ which was discussed in Sec. II B.

Let us now consider higher values of $m$. In general, we can go from
the states near $-\pi /2$ to states near $\pi /2$ in $m$ steps, either
through the intermediate states with increasing momenta at $k_j = -\pi /2 
+ j\pi /m$ with $j=1,2,...,m$, or through the intermediate states with 
decreasing momenta at $k_j = -\pi /2 - j\pi /m$ with $j=1,2,...,m$. In 
each case, we get a product of $m$ matrix elements from (\ref{matel1}) 
divided by $m-1$ energy factors like $E_j -E_F$ where $E_F =0$. To be 
explicit, this product is
\beq
F_m ~=~ (\delta /2)^m ~\frac{\prod_{p=0}^{m-1} \sin [(p+\frac{1}{2}) \pi 
/n]}{\prod_{p=1}^{m-1} \sin [p\pi /m]} ~=~ \frac{\delta^m}{m~2^m}.
\eeq
(The second equality follows from the identity
\beq
\prod_{p=1}^{n-1} ~[ 2 \sin (p\pi /n)] ~=~ \lim_{z \rightarrow 1} ~
\prod_{p=1}^{n-1} ~(z - e^{i2\pi p/n}) ~=~ \lim_{z \rightarrow 1} ~
\frac{z^n - 1}{z-1} ~=~ n ~).
\eeq
After adding the contributions from the two sets of intermediate states
and shifting $x$ appropriately, we get a term in the continuum 
Hamiltonian of the form 
\beq
2F_m ~(~ \psr^{\dagger} \psl ~+~ \psl^{\dagger} \psr ~)~ \cos (mqx) ~.
\eeq
This gives rise to a Dirac equation whose form is identical to that in Eq. 
(\ref{eom1}) except that $q$ and $\delta$ in that equation are now
replaced by $mq$ and 
\beq
2F_m ~=~ \frac{\delta^m}{m~2^{m-1}}
\label{fm}
\eeq
respectively. We therefore
get the same pattern of bands and gaps as in Figs. 1 and 2; we recall here
that those figures only show one-fourth of the region of interest since
they are restricted to positive values of $q$ and $E$. The complete regions
which resemble crosses are visible in Fig. 7 along the line $E=0$ at the 
values $Q/\pi = 1/m$. The change of scale from $\delta$ to $2F_m$ in 
Eq. (\ref{fm}) explains why the sizes of the crosses decrease rapidly as 
$m$ increases.

This concludes our discussion of the main features of the complete
energy spectrum with incommensurate hopping. Due to the close relationship 
of our work with the anisotropic AH problem, it is not surprising that
our Fig. 7 resembles the complete energy spectrum with an incommensurate 
on-site potential as given in, say, Fig. 7 of Sun and Ralston \cite{azb}.
(Their variables $\Phi /\Phi_0$ and $\lambda$ correspond to our $Q/(2\pi)$ 
and $1/\delta$ respectively). For instance, their figure also shows a 
series of crosses along the line $E=0$. Due to the differences in the 
matrix elements in (\ref{matel1}) versus the ones in (\ref{matel2}), the 
scales of the crosses go as
\beq
2F_m ~=~ \frac{\delta^m}{m}
\eeq
for the case of on-site incommensuration. This decreases more slowly 
than (\ref{fm}) with increasing $m$ which explains why the crosses in Sun 
and Ralston are bigger than the ones in our Fig. 7; they also have larger 
values of $\delta$ equal to $1$ and $2/3$ compared to our $\delta =0.4$. 
Similarly, their figure shows the smaller gaps on the edges of the 
larger gaps as explained by our case B.

\section{Outlook}

The continuum method discussed in this paper is robust in the sense that 
it can be applied even if we add weak next-nearest-neighbor hoppings or the
incommensuration is given by the sum of several harmonics. This is an
advantage over some other analytical methods. For instance, the Bethe
ansatz discussed by Wiegmann and Zabrodin \cite{azb} only works in somewhat
special cases; in those cases, however, the Bethe ansatz works for all values
of $\delta$. A limitation of the continuum method is that it is accurate 
only if the incommensuration is weak.

An interesting problem for future study may therefore 
be to examine what happens if the incommensuration in the hopping 
is strong, i.e., if $\delta$ is comparable to or larger
than $1$. [Both the perturbative and the continuum Dirac theory (or 
bosonization) approaches would then fail. Further, for strong 
incommensuration, it does make a difference whether the incommensurate term 
is in the hopping or on-site; the latter case has been studied much more
extensively in the literature]. The pattern of bands and gaps 
is expected to be much more complicated in that case, perhaps with a 
devil's staircase or a point spectrum structure \cite{sok,ost,lyu,azb}.
The nature of the wave functions, namely, whether they remain extended or
become localized, would also be of interest.

\vskip .7 true cm
\noindent {\bf Acknowledgments}

We thank G. Ananthakrishna, Somen Bhattacharjee, Rahul Pandit and, specially, 
Apoorva Patel for many useful discussions.

\newpage


\newpage
\vskip 1 true cm
\noindent {\bf Figure Captions}
\vskip .5 true cm

\noindent
{1.} The bands as a function of the energy $E$ and the wave number $q$ (near
dimerization), both in units of $\delta$, taking $\delta =0.05$. The 
numbers $1$, $3$ and $5$ labeling the three biggest gaps are explained in 
the text.

\noindent
{2.} A finer view of the bands as a function of $E/ \delta$ and $q/ \delta$,
for $\delta = 0.05$. A comparison with the second-order perturbation 
results is indicated by the six solid lines.

\noindent
{3.} Semi-log plot of the widths $\Delta E_n /\delta$ of the lowest four bands 
$n=0,1,2,3$. The points indicate the values obtained numerically for
$\delta = 0.05$, while the solid lines (labeled by $n$) show the WKB 
expressions in Eq. (\ref{wid1}).

\noindent
{4.} The density of states $\rho$ as a function of $E/\delta$ near 
dimerization. The solid and dashed lines show the behavior for $q \rightarrow 
0$ and $q=0$ respectively.

\noindent
{5.} The bands as a function of $(E-E_F)/\delta$ and $q/\delta$, for 
$Q_0 = 2\pi /3$ and $\delta =0.4$, with resolution $dE = 10^{-5}$. 

\noindent
{6.} Details of the bands as a function of $(E-E_F)/\delta$ and $q/\delta$, 
near the top of the $Q_0 = 2\pi /3$ gap for $\delta =0.4$, with resolution 
$dE = 10^{-8}$. 

\noindent
{7.} The bands as a function of $E$ and $Q/(2\pi)$ for $\delta =0.4$, with 
resolution $dE = 10^{-4}$. 

\newpage

\begin{figure}[ht]
\begin{center}
\epsfig{figure=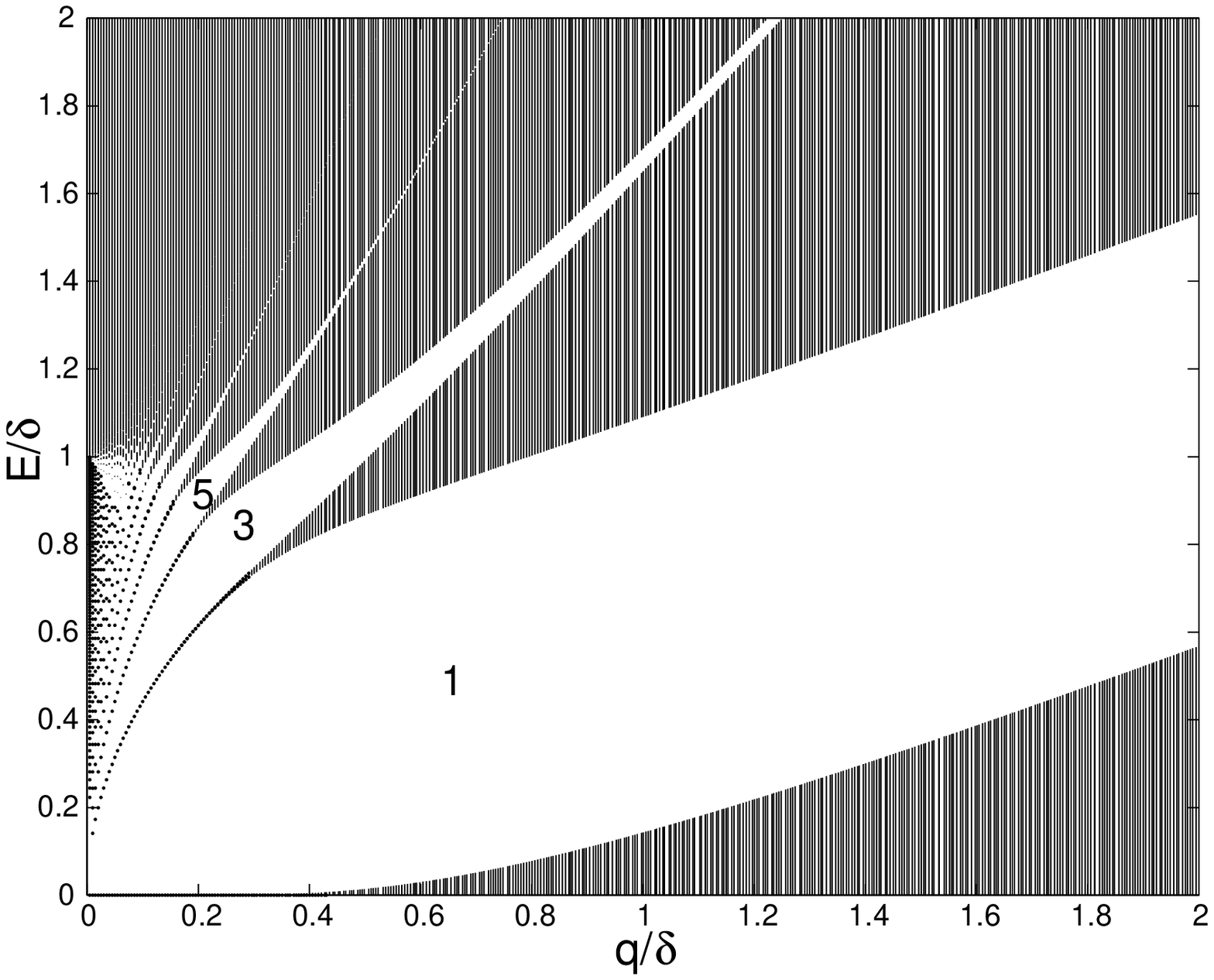,width=11cm}
\end{center}
\caption{}
\label{coarse}
\end{figure}

\vspace*{1cm}
\begin{figure}[hp]
\begin{center}
\epsfig{figure=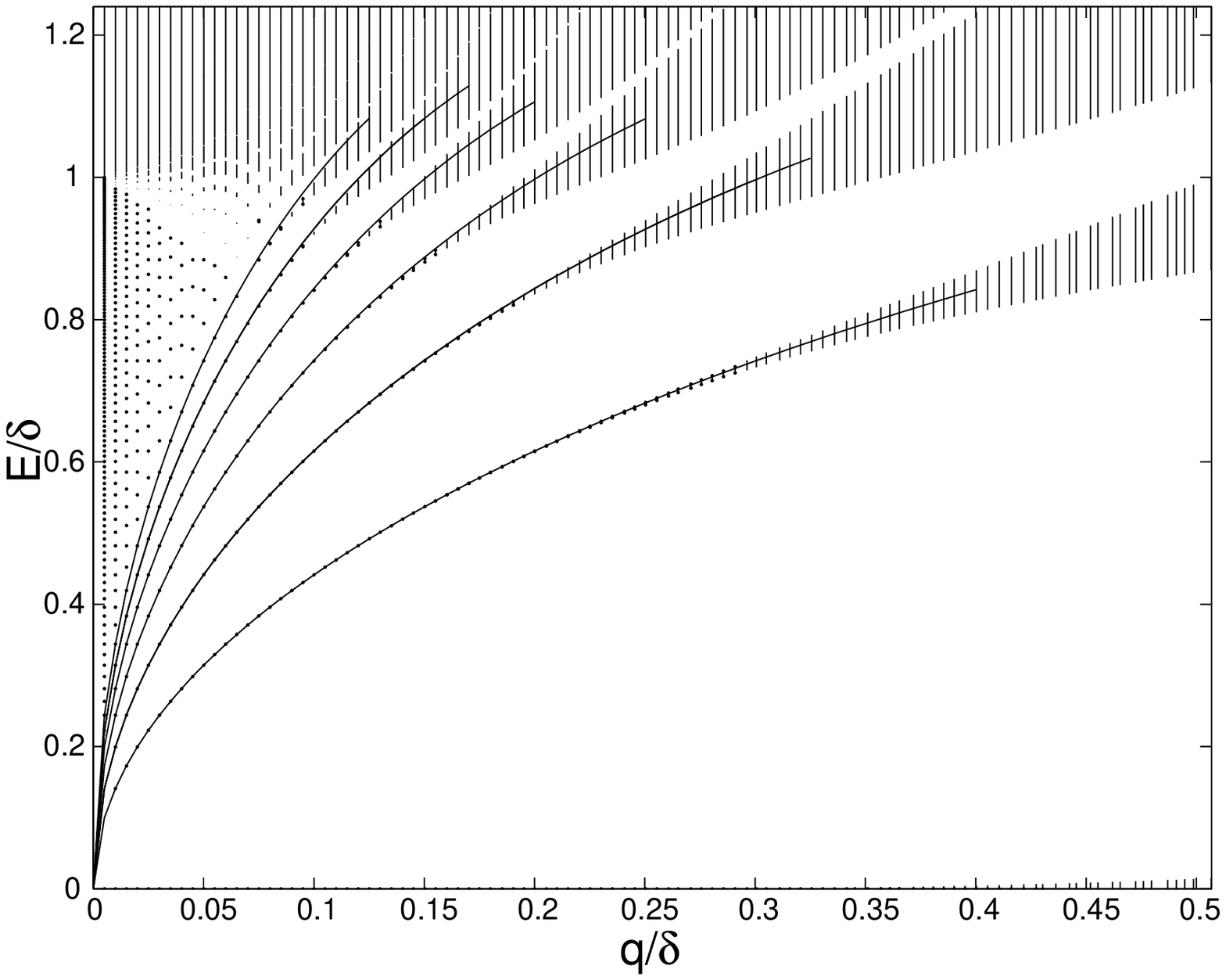,width=11cm}
\end{center}
\caption{}
\label{fine}
\end{figure}

\begin{figure}[ht]
\begin{center}
\epsfig{figure=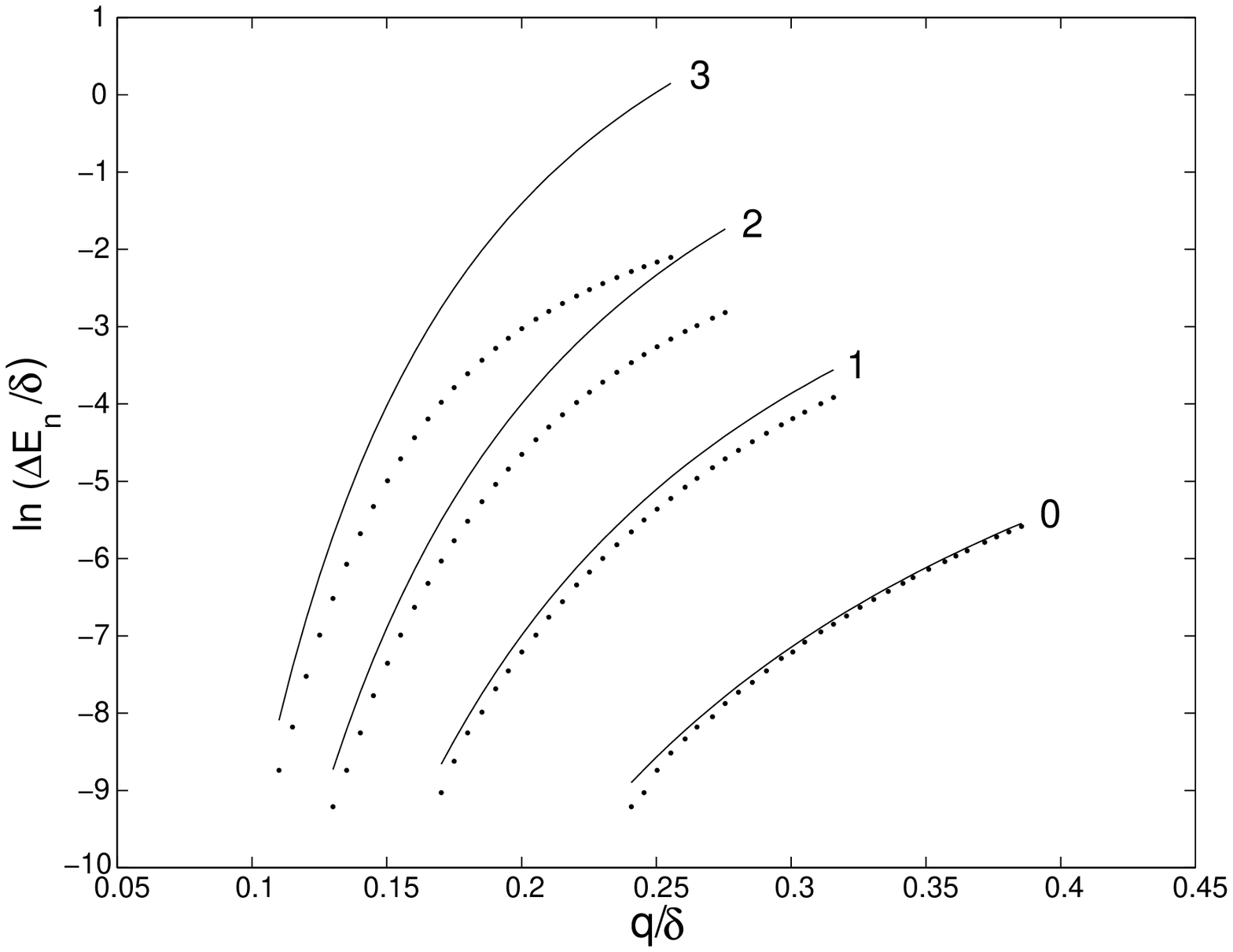,width=11cm}
\end{center}
\caption{}
\label{wkbfig}
\end{figure}

\vspace*{1cm}
\begin{figure}[hp]
\begin{center}
\epsfig{figure=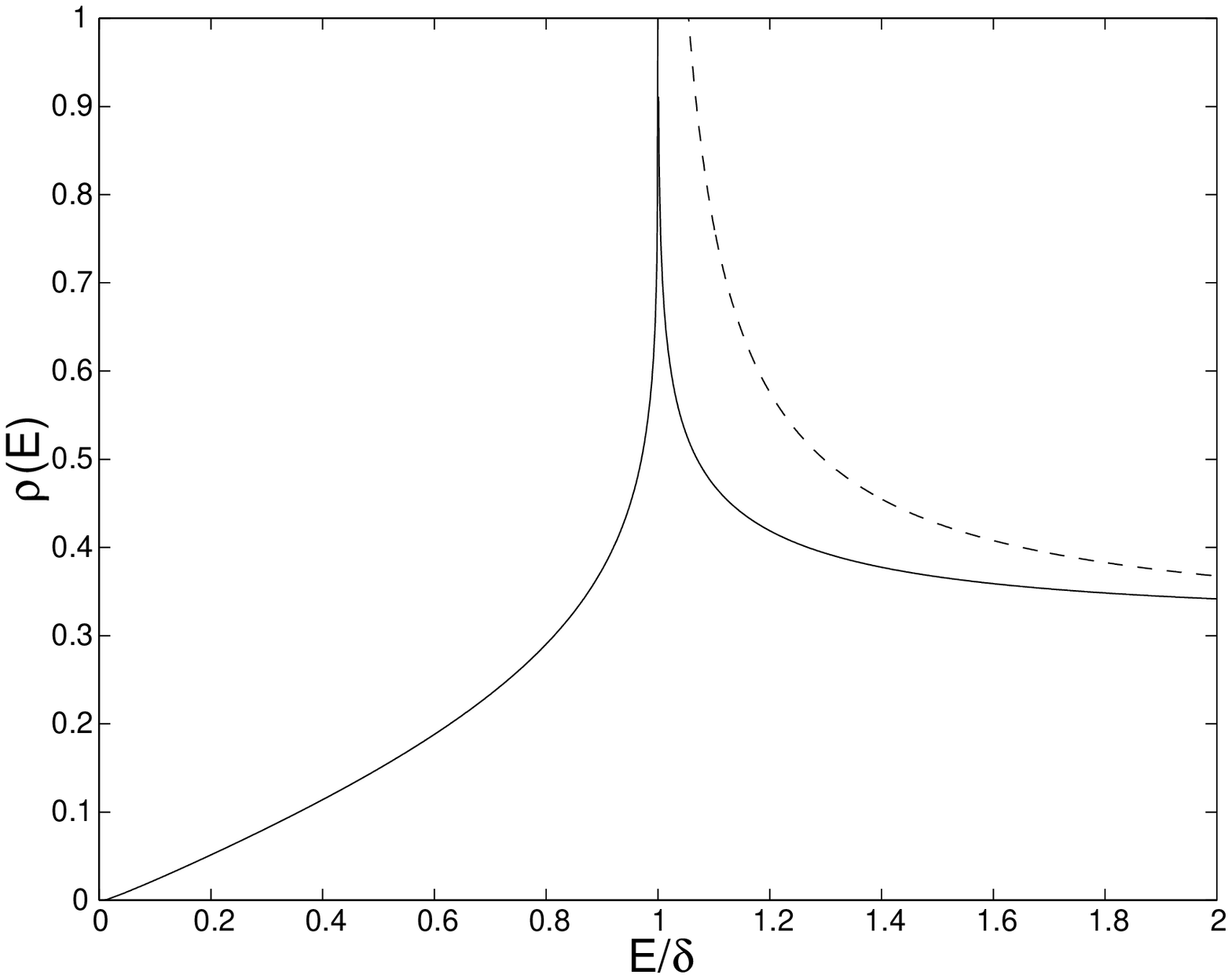,width=11cm}
\end{center}
\caption{}
\label{density}
\end{figure}

\begin{figure}[ht]
\begin{center}
\epsfig{figure=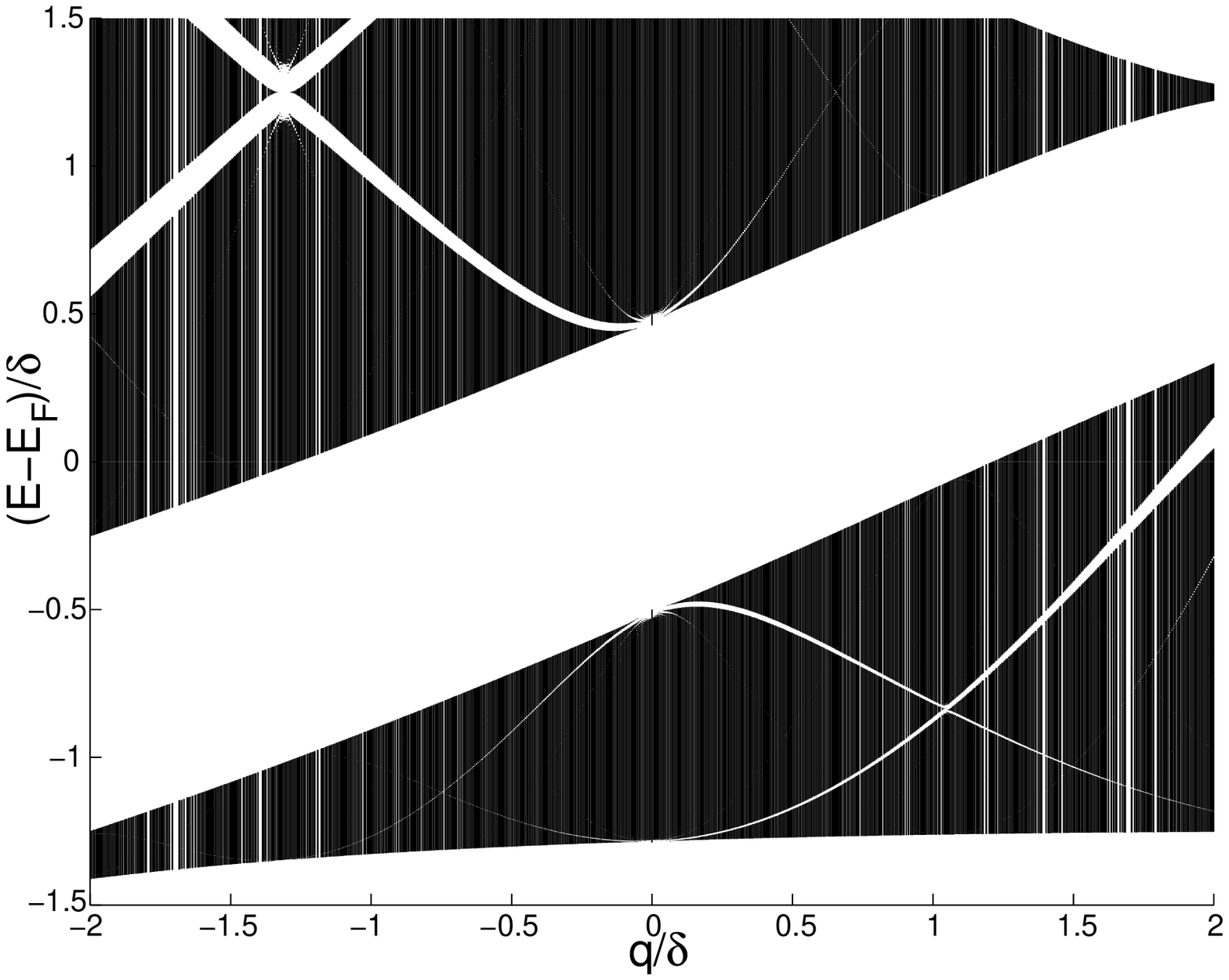,width=11cm}
\end{center}
\caption{}
\label{2pi3}
\end{figure}

\vspace*{1cm}
\begin{figure}[hp]
\begin{center}
\epsfig{figure=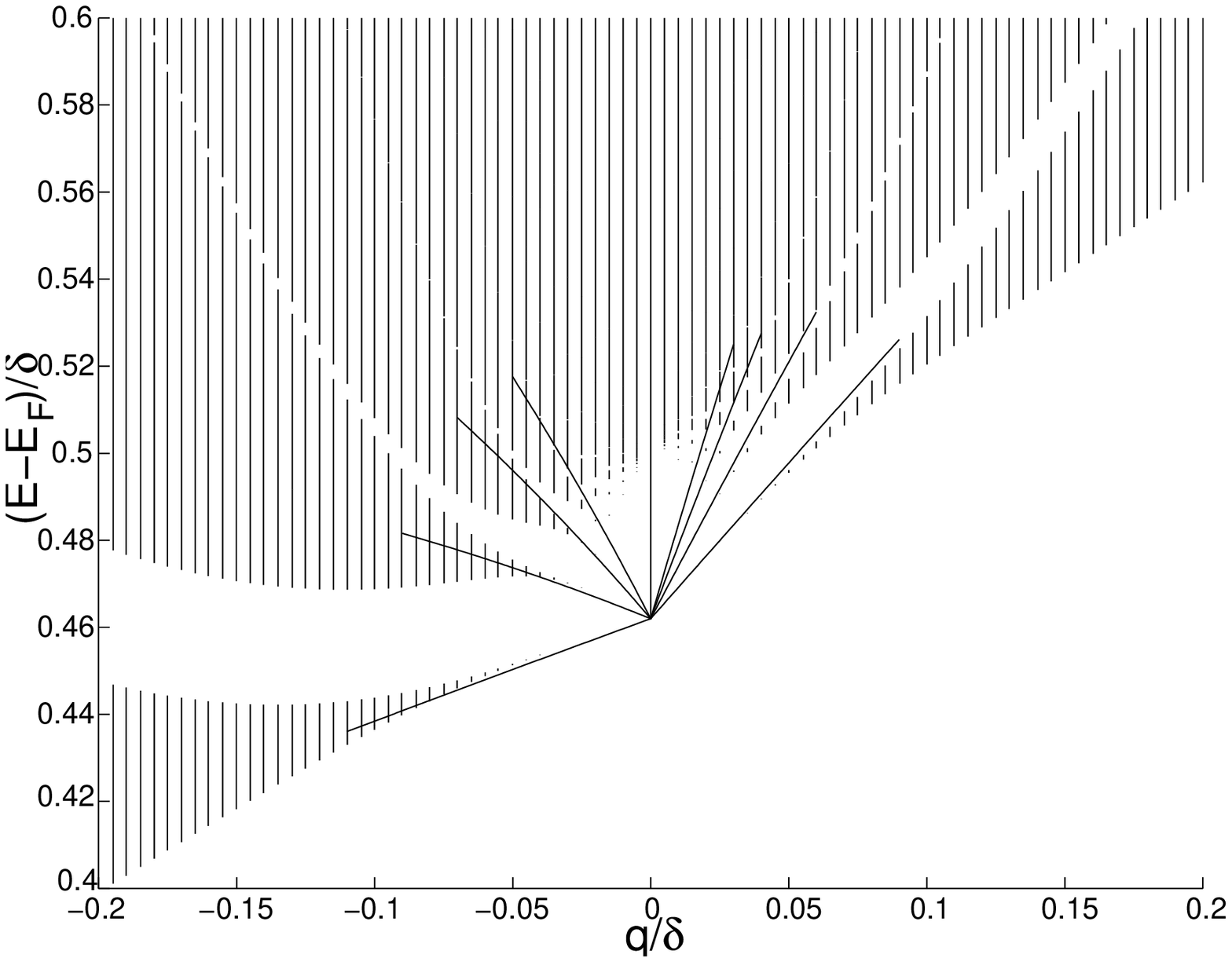,width=11cm}
\end{center}
\caption{}
\label{2pi3detail}
\end{figure}

\begin{figure}[ht]
\begin{center}
\epsfig{figure=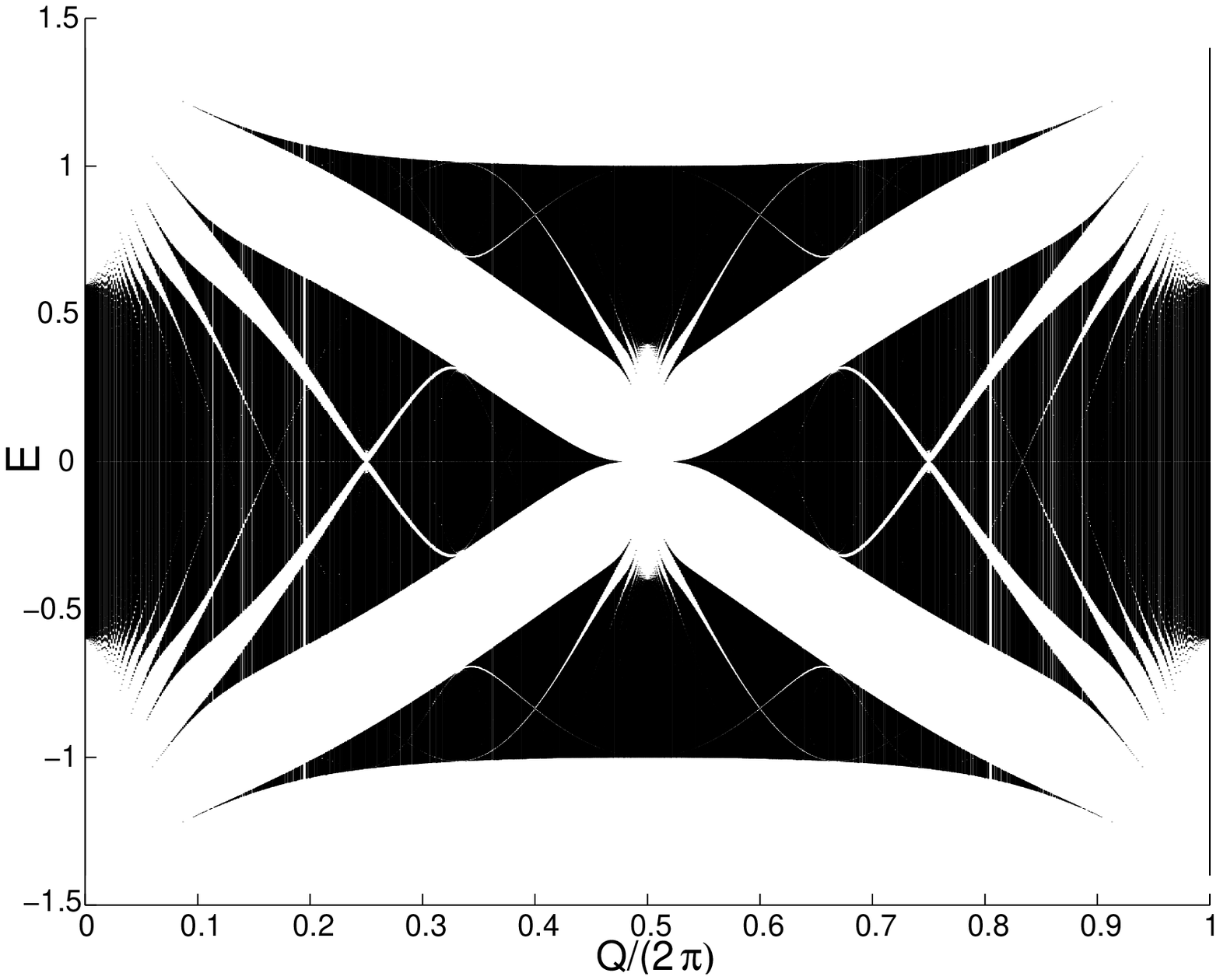,width=11cm}
\end{center}
\caption{}
\label{total}
\end{figure}

\end{document}